\RequirePackage{fix-cm}
\documentclass[smallextended]{svjour3}       
\smartqed  
\usepackage{graphicx}
\usepackage{amsfonts}
\newcommand\Half{{\frac{1}{2}}}
\newcommand\scriptHalf{{\scriptstyle\frac{\scriptstyle 1}{\scriptstyle 2}}}
\newcommand\Intd{{\mathrm{d}}}
\newcommand\Remove[1]{{\raise 1.2ex\hbox{$\times$}\kern-0.8em \lower 0.35ex\hbox{$#1$}}}
\newcommand\SmallFrac[2]{{\scriptstyle\frac{\scriptstyle #1}{\scriptstyle#2}}}

\newcommand\VEV[1]{{\left<0\right|#1\!\left|0\right>}}
\newcommand\eqdef{{\stackrel{\scriptstyle\mathrm{def}}{=}}}
\newcommand\rme{{\mathrm{e}}}
\newcommand\rmi{{\mathrm{i}}}

\newcommand{\uu}{{\mathfrak{u}}}
\newcommand{\emF}{{\hat\mathbf{F}}}\newcommand{\emA}{{\hat\mathbf{A}}}\newcommand{\emJ}{{\hat\mathbf{J}}}
\newcommand{\fkf}{{\mathfrak{f}}}\newcommand{\fkg}{{\mathfrak{g}}}
\newcommand\iS{{\rmi\hspace{-0.2ex}\mathsf{S}}}
\newcommand\iD{{\rmi\hspace{-0.2ex}\Delta}}
\newcommand{\bOP}{{\hat\mathsf{b}}}\newcommand{\dOP}{{\hat\mathsf{d}}}

\newcommand\Vip[2]{{#1\!\cdot\!#2}}
\newcommand\sVip[2]{{#1\cdot#2}}
\newcommand\kInt[1]{{\frac{\Intd^4#1}{(2\pi)^4}}}

\journalname{Foundations of Physics}
\begin{document}

\title{Nonlinear dependence of the renormalization scale on test functions}
\author{Peter Morgan}
\institute{\emph{Present address:}Physics Department, Yale University, New Haven, CT 06520, USA.
           \email{peter.w.morgan@yale.edu}}

\date{Received: date / Accepted: date}
\maketitle

\begin{abstract}
Quantum electrodynamics exhibits an informal nonlinear dependence on Lorentz invariant test function properties that
determine the renormalization scale, such as Mandelstam variables, contrary to the linear dependence on test functions
that is required by the Wightman axioms.
A first example of an alternative interacting quantum field formalism that has a comparable weakly nonlinear dependence
on Dirac spinor test functions is constructed, using $U(1)$--gauge connections and $U(1)$--gauge invariant
Dirac spinor test functions.
\end{abstract}

\section{Introduction}\label{introduction}
Gauge invariance and covariance are significant guides in the construction of quantum field theories (QFTs) in Lagrangian or
Hamiltonian formalisms.
Here, instead of introducing nonlinear modifications of the free quantum field Lagrangian or Hamiltonian dynamics and the
accompanying renormalization, the Lorentz invariant $U(1)$--gauge invariance that is required for observables of quantum
electrodynamics is ensured by introducing a nonlinear construction of $U(1)$--gauge connections and $U(1)$--invariant Dirac
spinor test functions in Section \ref{GTDS}, which may be used as part of a $U(1)$--gauge invariant formalism to construct
a wide range of interacting models.
We thereby avoid the introduction of products of distributions and, because only direct space--time relationships between
test functions contribute to Vacuum Expectation Values (VEVs), we avoid divergent integrations over intermediate loop momenta.
As a consequence of this formalism, observables become weakly nonlinear functionals of the test functions that are used to model
the various wave packets and measurement windows that are used in an experimental apparatus, so that the quantum fields
will no longer be operator--valued distributions.

Section \ref{NLQF} compares the weakly nonlinear dependence on test functions that is introduced here with the nonlinear
dependence on experimental scales that is introduced by the Lorentz invariant choice of a renormalization scale in
Lagrangian and Hamiltonian formalisms to interacting QFT.
It is pointed out that the renormalization scale is a Lorentz invariant function of properties of the set of wave packets
that are prepared and measured in an experiment, informally dependent on wave--number properties such as the Mandelstam
variables, so that Lagrangian and Hamiltonian formalisms implicitly use nonlinear functionals of test functions.
The nonlinear dependence is relatively weak, however, insofar as the renormalization scale is independent of the amplitudes
of the test functions, so that homogeneity (of degree 1) is preserved.

The reconciliation of the mathematically coherent Wightman axioms with the empirically successful Lagrangian and
related formalisms to interacting QFT is a longstanding problem~\cite{Fredenhagen2007,Buchholz2000}.
In an attempt to make the Wightman axioms more applicable, we will here relax the postulate that a Wightman field
$\hat\phi(x)$ is an operator--valued \emph{distribution}, whereby the operators of the theory are obtained by
linear ``smearing'', for a well--behaved linear space $\mathcal{S}$ of ``test'' functions,
\begin{equation}\label{QFdefinition}
  \hat\phi:\mathcal{S}\rightarrow\mathcal{A};f\mapsto\hat\phi_f= \int\hat\phi(x)f(x)\Intd^4x,
\end{equation}
which has hitherto been little addressed in the literature.
Our point of departure will be to take a map $\hat\xi:f\mapsto\hat\xi_f$, from a suitably well--behaved space of test functions
into a $*$-algebra of operators, to be a weakly nonlinear functional of the test functions, so that in general
$\hat\xi_{f+g}\not=\hat\xi_f+\hat\xi_g$, essentially denying the linearity that is implied by Eq. (\ref{QFdefinition}),
however we will continue to require homogeneity of degree 1 over the reals, $\hat\xi_{\lambda f}=\lambda\hat\xi_f$.
Nonetheless, the action of the $*$-algebra of operators on Hilbert space vectors is taken to be linear,
$(\hat\xi_f+\hat\xi_g)\left|\psi\right>=\hat\xi_f\left|\psi\right>+\hat\xi_g\left|\psi\right>$ and
$(\lambda\hat\xi_f)\left|\psi\right>=\lambda(\hat\xi_f\left|\psi\right>)$, and we will retain the usual Born rule
construction of expected values, of probabilities, and of correlations that is common to all quantum theory ---and their
use to model the statistics of multiple experimental datasets--- and the converse Gelfand--Naimark--Segal (GNS)--construction
of a Hilbert space from the expected values that are generated by a state over a $*$-algebra of operators~\cite[\S III.2]{Haag}.
This ``Born--GNS'' linearity is effective, as in Lagrangian QFT, as quantum theory's way of generating probabilities that
satisfy the Kolmogorov axioms of probability, however the linearity we relinquish has no comparable necessity.

\section{Nonlinearity and renormalization}\label{NLQF}
We first note that real--space renormalization quite strongly motivates some kind of nonlinear structure.
As a loose description of the general process, we have from Wilson and Kogut,
\begin{quote}
    ``The [first] basic idea [of the renormalization group] is the same as in hydrodynamics.
    In hydrodynamics one introduces new variables such as the density $\rho(x)$ which represents an average
    over the original microscopic degrees of freedom.''~\cite[p. 79]{WilsonKogut}
\end{quote}
The processes that are used to construct higher--level block variables, however, such as \emph{majority rule} and \emph{decimation},
are not in general expressible as linearly constructed weighted averages of lower--level block variables.
A full description of the process of real--space renormalization might be a list of nonlinear renormalization functionals, in
which successive levels are constructed from the level below, so that with explicit indexing we might write
$\rho_{[F_1,F_2,...]}(x)$ instead of $\rho(x)$ (although networks of renormalizations are also possible), but
a minimal systematic formalism is to use a mass scale $\mu$ to generate a list of nonlinear renormalization functionals that
allows us to construct sequences $\rho_{\mu_1}(x),\rho_{\mu_2}(x),...$.
We may, for example, introduce a single nonlinear functional $F$ to construct $F:\rho_{\mu_i}\mapsto\rho_{\mu_{i+1}}$,
with $\mu_{i+1}<\mu_i$ because renormalization acts as a semi--group that only transforms towards coarser descriptions at
lower mass scales, or, as a continuous construction, we may introduce a parameterized renormalization functional such as
$F_\lambda:\rho_{\mu}\mapsto\rho_{\lambda\mu}$, $\lambda<1$.
For such a construction, $\mu$ parameterizes the nonlinear renormalization process, so that the phenomenology associated with
a given $\rho_\mu(x)$ depends nonlinearly on $\mu$.
A more recent account makes explicit the dependence of the dynamics on scale, found everywhere in the literature and in textbooks,
\begin{quote}
    ``the natural description of physics generally changes with the scale at which observations are made.
    Crudely speaking, this is no more high--minded a statement than saying that the world around us looks
    rather different when viewed through a microscope.
    More precisely, our parametrization of some system in terms of both the degrees of freedom and an action
    specifying how they interact generally change with scale.''~\cite[p. 178]{Rosten}
\end{quote}
Both for real--space renormalization and for Lagrangian and Hamiltonian formalisms, the renormalization scale determines the
dynamical constants of an effective (quantum) field theory, both the effective masses of the fields and the strengths
of the interactions between them.

Less minimally than a one--dimensional scale $\mu$, we can adapt the use of a test function $w(x)$ in quantum field theory
in a linear convolution with an operator--valued distribution $\hat\phi(x)$ to obtain a linear space of quantum field
operators $\hat\phi_w(x)=[\hat\phi\star w](x)=\int\hat\phi(y)w(x-y)\Intd^4y$, by interpreting the test function $w(x)$
as an infinite--dimensional description of the many scales at which a given measurement or preparation operates, with
varying amplitudes at different wave--numbers.
If it is taken that the test function is a parameterization of a nonlinear renormalization process rather than of a linear
smearing, it is natural to suppose that the phenomenology associated with a given $\rho_w(x)$ will in general depend
Lorentz covariantly and nonlinearly on $w(x)$, although our experience clearly suggests that the nonlinearity is
relatively weak.

Renormalization in Lagrangian and Hamiltonian formalisms ordinarily makes no formal connection between the renormalization
scale that is used in calculations for a given interacting quantum field theory and the scales that are implicitly
introduced by the test functions that model particular measurements and preparations, but there is a clear pragmatic
connection, in that the highest rest--mass that is used in each test function and Lorentz invariant experimental
parameters such as Mandelstam variables that are determined by the relationships between several test functions
all contribute to the determination of the renormalization scale to be used.
Indeed, because the various masses associated with the dynamics are all determined by the renormalization scale,
Lorentz invariant properties of the test functions that describe the various wave packets and of the relationships between
the test functions are the only available guides for the renormalization scale, unless we take the renormalization scale to
be entirely independent of the test functions that describe the experimental apparatus in detail.

Although there is a Lorentz invariant dependence of the renormalization scale on the wave--numbers that are used in test
functions, the renormalization scale is generally independent of the amplitudes of the test functions.
At least in the first instance, therefore, it seems a worthwhile mathematical simplification to require models to be
homogeneous of degree 1 over the reals in the test functions.

An advantage of introducing a weak nonlinear functional dependence on detailed Lorentz invariant properties of all test functions,
as in Section \ref{GTDS}, is that it allows widely varying measurement scales of the different measurement and preparation apparatuses
that are part of a collection of experiments to be accommodated naturally, where a single renormalization scale would only inadequately
reflect the complexity of the experiments.
Insofar as more sophisticated experiments introduce dependencies on multiple length--scales, it may become necessary for more
sophisticated models of the nonlinear dependence on experimental contexts to be used than the current dependence on a single
renormalization scale, and, conversely, this theoretical focus requires an experimental focus on the effects of introducing
or discovering dependencies on multiple length--scales.

\section{$U(1)$--gauge transformations for Dirac spinor test functions}\label{GTDS}
If the Dirac spinor operator--valued distribution $\hat\psi(x)$ gauge transforms as
$\hat\psi(x)\mapsto\rme^{\rmi\theta(x)}\hat\psi(x)$, with $\theta(x)\in\mathbb{R}$, a natural choice is to take a
corresponding Dirac spinor test function $U(x)$ to gauge transform contravariantly as $U(x)\mapsto\rme^{-\rmi\theta(x)}U(x)$,
so that the scalar operator
$$\hat\psi^{\ }_U=\int \overline{U^c(x)}\hat\psi(x)\Intd^4x\mapsto\int \overline{U^c(x)}\hat\psi(x)\Intd^4x$$
is $U(1)$--gauge invariant (where the charge conjugate $U^c(x)$ of $U(x)$ is used to ensure that $\hat\psi^{\ }_U$
follows the convention that it should be complex--linear in $U$; Appendix \ref{DiracAppendix} discusses
the relation of this to the conventional notation).

The usual anti--commutator and two--point VEVs,
\begin{eqnarray*}
  \{\hat\psi_U^{\ },\hat\psi_V^{\;\dagger}\}=
      (V,U)&=&\!\int\!\frac{\Intd^4k}{(2\pi)^4}2\pi\delta(\Vip{k}{k}-m^2)
                                   \varepsilon(k_0)\overline{\tilde V(k)}(k_\mu\gamma^\mu+m)\tilde U(k),\cr
  \VEV{\hat\psi_U^{\ }\hat\psi_V^{\;\dagger}}=
    (V,U)_+&=&\!\int\!\frac{\Intd^4k}{(2\pi)^4}2\pi\delta(\Vip{k}{k}-m^2)
                                        \theta(k_0)\overline{\tilde V(k)}(k_\mu\gamma^\mu+m)\tilde V(k),
\end{eqnarray*}
however, are not invariant under the $U(1)$--gauge transformation of $U(x)$ and $V(x)$.
To modify these expressions to be $U(1)$--gauge invariant, we introduce, as a first and most natural example, a Lorentz covariant
$U(1)$--gauge connection that is constructed as a local functional of the Dirac spinor test function $U(x)$,
$$\uu_\alpha[U](x)=\frac{\rmi}{2}
      \left[\frac{\overline{U(x)}\gamma^\mu U(x)\overline{U(x)}\gamma_\mu\partial_\alpha U(x)}
           {\overline{U(x)}\gamma^\mu U(x)\overline{U(x)}\gamma_\mu U(x)}
           -\mathrm{c.c.}\right],
$$
which $U(1)$--gauge transforms, when $U(x) \mapsto \rme^{-\rmi\theta(x)}U(x)$, as
$$\uu_\alpha[U](x) \mapsto \uu_\alpha[U](x)+\partial_\alpha\theta(x);$$
a class of such connections is constructed in Appendix \ref{GaugeConnections}.
Using this $U(1)$--gauge connection, we may apply a standard method for constructing a $U(1)$--gauge invariant object,
\begin{eqnarray*}
  U_\uu(x)&=&\exp{\!\left[\rmi\!\int\!\Xi^\mu(x-y)\uu_\mu[U](y)\mathrm{d}^4y\right]}U(x),\\
  U_\uu(x)&\mapsto&\exp{\!\left[\rmi\!\int\!\Xi^\mu(x-y)\partial_\mu\theta(y)\mathrm{d}^4y\right]}
                 \rme^{-\rmi\theta(x)}U_\uu(x)\\
        &=&\exp{\!\left[\rmi\!\int\!\partial_\mu\Xi^\mu(x-y)\theta(y)\mathrm{d}^4y\right]}
                 \rme^{-\rmi\theta(x)}U_\uu(x)=U_\uu(x),
\end{eqnarray*}
where we require that $\theta(x)$ diminishes sufficiently fast at infinity and that the 4--vector distribution
$\Xi^\mu(x)$ satisfies $\partial_\mu\Xi^\mu(x)=\delta^4(x)$ in a way that ensures Lorentz and translation invariance
of the construction of $U_\uu$.
One example that ensures this is for $\Xi^\mu(x)$ to be the gradient of a zero--mass Green's function in 3+1--dimensions,
such as $\Xi^\mu(x)=\partial^\mu G_{\mathrm{ret}}(x)$.
For this case, it is sufficient for $U_\uu(x)=U(x)$ if $\uu_\mu[U](x)$ is divergence--free, $\partial^\mu\uu_\mu[U](x)=0$,
and $\uu_\mu[U](x)$ decreases sufficiently rapidly at infinity; we find that the test function $U_\uu(x)$ satisfies this
condition, because
\[
  \uu_\alpha[U_\uu](x)=\uu_\alpha[U](x)-\scriptHalf\left[\int\partial_\alpha\partial^\mu G_{\mathrm{ret}}(x-y)
                                                          \uu_\mu[U](y)\Intd^4y+\mathrm{c.c.}\right],
\]
so that
\[
  \partial^\alpha\uu_\alpha[U_\uu](x)=\partial^\alpha\uu_\alpha[U](x)
               -\scriptHalf\left[\int\partial^\mu\delta^4(x-y)\uu_\mu[U](y)\Intd^4y+\mathrm{c.c.}\right]=0,
\]
and, consequently, $(U_\uu)_\uu(x)=U_\uu(x)$ is for this case idempotent.
For this case, we may regard the operation $U\mapsto U_\uu$ as a way to construct a test function for which the chosen connection
$\uu_\alpha[U_\uu]$ is divergence--free; how the operation achieves the divergence--free property is fixed by the choice of
$\Xi^\mu$.
Other examples are constructed in Appendix \ref{GaugePropagators}, for some of which $\Xi^\mu(x)$ has components that are
divergences of Dirac algebra--valued bivector fields as well as components that are gradients of scalar fields.

Requiring that the positive semi--definite inner products $(U_\uu,U_\uu)$ and $(U_\uu,U_\uu)_+$ exist is taken to be
a nontrivial defining constraint on the Dirac spinor test function space.
For example, for the ansatz $U(x)=E(x)\rme^{-\rmi\sVip{k}{x}\Phi(x)}U_0$, where $E(x)$ is a real--valued positive--definite
envelope, $\Phi(x)$ is a real--valued long--range regularization, and $U_0$ is a constant Dirac spinor,
$\uu_\alpha[U](x)=\partial_\alpha[\Vip{k}{x}\Phi(x)]$, independent of both $E(x)$ and $U_0$; it is enough for $(U_\uu,U_\uu)$
and $(U_\uu,U_\uu)_+$ to exist for this case if $E(x)$ and $\Phi(x)$ are smooth and decrease rapidly for large $x$, whereas
for the quantized free Dirac spinor field $\Phi(x)$ may be constant.
For this simplest case, $U_\uu(x)=E(x)U_0$ is independent of $k$ and of $\Phi(x)$, with only the constant relative
phases of the components of the constant spinor $U_0$ surviving.
For less straightforward cases, however, $\uu[U](x)$ will in general not be a derivative of a scalar, so that the curvature
bivector $\Intd\uu_{\mu\alpha}[U]=\partial_{[\mu}\uu_{\alpha]}[U]$ will in general not be trivial and $U_\uu(x)$ will have
surviving nontrivial variations of relative phases of its components.
For example, for constant Dirac spinors $u^{(1)}$ and $u^{(2)}$, representing spin--up and spin--down in some frame, for
which $\overline{u^{(1)}}u^{(1)}=\overline{u^{(2)}}u^{(2)}=1$,
$\overline{u^{(1)}}u^{(2)}=0=\overline{u^{(1)}}\rmi\gamma^5 u^{(1)}=
 \overline{u^{(1)}}\rmi\gamma^5 u^{(2)}=\overline{u^{(2)}}\rmi\gamma^5 u^{(2)}$, for the Dirac spinor test function
\begin{equation}\label{FFexample}
  U(x)=E_1(x)\rme^{-\rmi\sVip{k_1}{x}\Phi_1(x)}u^{(1)}+E_2(x)\rme^{-\rmi\sVip{k_2}{x}\Phi_2(x)}u^{(2)},
\end{equation}
for which $\overline{U(x)}\gamma^\mu U(x)\overline{U(x)}\gamma_\mu U(x)=
[\overline{U(x)}U(x)]^2+[\overline{U(x)}\rmi\gamma^5 U(x)]^2$ reduces to $\Bigl(E_1(x)^2+E_2(x)^2\Bigr)^2$,
we obtain
\begin{eqnarray*}
  \uu_\alpha[U]&=&\frac{E_1(x)^2\partial_\alpha[\Vip{k_1}{x}\Phi_1(x)]+E_2(x)^2\partial_\alpha[\Vip{k_2}{x}\Phi_2(x)]}
                       {E_1(x)^2+E_2(x)^2},\cr
  \Intd\uu_{\mu\alpha}[U]&=&\partial_{[\mu}\!\left[\frac{E_1(x)^2}{E_1(x)^2+E_2(x)^2}\right]\!\partial_{\alpha]}[\Vip{k_1}{x}\Phi_1(x)]\cr
         &&\hspace{2em}    +\partial_{[\mu}\!\left[\frac{E_2(x)^2}{E_1(x)^2+E_2(x)^2}\right]\!\partial_{\alpha]}[\Vip{k_2}{x}\Phi_2(x)].
\end{eqnarray*}
If over some large region we take $\Phi_1(x)\approx 1$ and $\Phi_2(x)\approx 1$ to be constant, $E_1(x)\approx\cos(\Vip{k_3}{x}+\theta)$,
and $E_2(x)\approx\sin(\Vip{k_3}{x}+\theta)$, then in that region we obtain
$\overline{U(x)}\gamma^\mu U(x)\overline{U(x)}\gamma_\mu U(x)\approx 1$ and
\begin{eqnarray*}
  \uu_\alpha[U]&\approx&\cos^2(\Vip{k_3}{x}+\theta)k_{1\alpha}+\sin^2(\Vip{k_3}{x}+\theta)k_{2\alpha},\\
  \Intd\uu_{\mu\alpha}[U]&\approx& \sin(2\Vip{k_3}{x}+2\theta)k_{3[\mu}(k_2-k_1)_{\alpha]},
\end{eqnarray*}
so that for this ansatz the $U(1)$--gauge connection, with this gauge--fixing, oscillates between the spin--up wave--number $k_1$
and the spin--down wave--number $k_2$, following the doubled spin--up/spin--down wave--number $k_3$ that is associated with $E_1$
and $E_2$.
The $U(1)$--gauge curvature follows the same doubled spin--up/spin--down wave--number $k_3$, with amplitude determined by
the bivector generated by the spin--up/spin--down wave--number and the $U(1)$--gauge invariant difference between the spin--up
wave--number and the spin--down wave--number.
It is enough for $U_\uu$ to exist for this ansatz if $E_1(x)$ and $E_2(x)$ are smooth and $\Phi_1(x)$ and $\Phi_2(x)$ are smooth
and decrease rapidly at long range; it is enough for $(U_\uu,U_\uu)$ and $(U_\uu,U_\uu)_+$ to exist if $E_1(x)$ and $E_2(x)$ also
decrease rapidly at long range.
We note that although the construction of $U_\uu$ is mathematically straightforward it is not straightforward to write down
$U_\uu$ even for these elementary ans\"atze.

We also require that the $U(1)$--gauge invariant positive semi--definite inner product $(\Intd\uu[U],\Intd\uu[U])_0$ exists,
where $(\cdot,\cdot)^{\ }_0$ is the two--point VEV of the quantized free Maxwell bivector field,
$$\VEV{\emF_{\!\fkf}^\dagger\emF_{\!\fkg}^{\ }}=(\fkf,\fkg)^{\ }_0
         =-\int\!\frac{\Intd^4k}{(2\pi)^4}2\pi\delta(\Vip{k}{k})\theta(k_0)k^\alpha g^{\mu\nu} k^\beta
                 \tilde \fkf_{[\alpha\mu]}^*(k)\tilde \fkg_{[\beta\nu]}^{\ }(k).$$
As well as taking $U_\uu$ to be a test function for the quantized Dirac spinor field, $\Intd\uu[U]$ will also be taken to be a
test function for the quantized Maxwell bivector field, which, taken with other test functions, describe how a state preparation
modulates the vacuum state or how a measurement responds to a given prepared state; $\hat\psi(x)$ and $\emF(x)$ unmediated by
experimentally descriptive explicit choices of test functions are taken not to be accessible to experiment.
In the figurative language of signal analysis, a test function is referred to as a ``window function'', as if when making
a measurement we must say what kind of distorting window we are looking through; as for all test functions, the bivector
$\Intd\uu[U]$, for example, is \emph{not} the Maxwell bivector field, it is the window that mediates when we prepare or
examine the Maxwell bivector field using $\emF_{\!\Intd\uu[U]}$.
In terms of differential forms, in the absence of boundary terms, we could instead write $\emF_{\!\Intd\uu[U]}$ as
$\emA_{\delta\Intd\uu[U]}$, because $\emF=\Intd\emA$, or, because $\delta\emF=\emJ$,
\[\emJ_{\uu[U]}=\delta\emF_{\!\uu[U]}=\emF_{\!\Intd\uu[U]}=\Intd\emA_{\Intd\uu[U]}=\emA_{\delta\Intd\uu[U]},
\]
so that $\uu[U]$ is a 4--current test function (or, in fact, a 4--current test connection), and a given choice for $\uu[U]$
determines a constitutive relation for a given type of quantized Dirac spinor field, insofar as a given Dirac spinor test
function $U$ determines a 4--current test connection $\uu[U]$, and hence also a Maxwell bivector test function $\Intd\uu[u]$.
To emphasize again the vagaries of the duality between quantum fields as operator--valued distributions and test functions,
$\delta\Intd\uu[U]$ is a test function, not a test connection, for the electromagnetic field connection $\emA$, and
$\emA_{\delta\Intd\uu[U]}=\emF_{\!\Intd\uu[U]}$ is gauge invariant.
As well as the alternative terminology ``window function'' for ``test function'', a test function that is used to construct
a time--ordered VEV generating functional is said to be a ``source field''.

Using the above, we can now construct $U(1)$--gauge invariant anti--commutation relations and two--point VEVs for a quantized
Dirac spinor field as an elementary nonlinear deformation of the quantized free Dirac spinor field that preserves homogeneity,
\[  \hat\xi^{\ }_U=\hat\psi^{\ }_{U_\uu},
    \qquad\{\hat\xi^{\ }_U,\hat\xi_V^{\;\dagger}\!\}=(V^{\ }_\uu,U^{\ }_\uu),
    \qquad\VEV{\hat\xi_V^{\;\dagger}\hat\xi^{\ }_U\!}=(V^{\ }_\uu,U^{\ }_\uu)_+.
\]
Also naturally, as a phase shift that depends on what other Dirac spinor test functions we use to modulate the vacuum or as
simultaneous measurements, we may construct field operators that introduce a directly electromagnetic effect, for example,
$$\hat\Psi^{\ }_U=\rme^{\rmi\lambda\emF_{\!\Intd\uu[U]}}\hat\xi^{\ }_U,$$
so that a state preparation that uses this operator results in a measurable Maxwell field as well as a
measurable Dirac spinor field.
The use of the Dirac spinor test function $U$ in two nonlinear forms, as the $U(1)$--gauge invariant Dirac spinor test function
$U_\uu$ and as the $U(1)$--gauge invariant Maxwell test function $\Intd\uu[U]$, introduces a nontrivial relationship between the
quantized Dirac and Maxwell fields.
With this construction, the VEVs are $U(1)$--gauge invariant, but the operators $\hat\Psi^{\ }_U$ and $\hat\Psi^{\,\dagger}_U$
that are used to generate the VEVS are not.

The constructions given for $\hat\Psi^{\ }_U$ ensure homogeneity of degree 1 over the reals,
$\hat\Psi^{\ }_{\lambda U}=\lambda\hat\Psi^{\ }_U$, because $\uu[\lambda U]=\uu[U]$, where linearity in general is not
satisfied, $\hat\Psi^{\ }_{U+V}\not=\hat\Psi^{\ }_U+\hat\Psi^{\ }_V$.
[More generally, $\uu[U]$ is invariant under multiplication by a real--valued positive--definite scalar function $E(x)$,
$\uu[EU]=\uu[U]$.]
There is, unfortunately, no obvious empirically principled constraint that would require that the mathematically
natural exponential $\rme^{\rmi\lambda\emF_{\!\Intd\uu[U]}}$ should not be replaced by an arbitrary polynomial or special
function in $\emF_{\!\Intd\uu[U]}$, although tractability and the naturalness of a phase shift in this context are
perhaps good mathematical reasons.
With the introduction of a collection of different connections $\uu_n[U]$, we might also construct significantly less
tractable operators, such as $\sum_n\lambda_n(\emF_{\!\Intd\uu_n[U]})^n\hat\psi^{\ }_{U_{\uu_n}}$, which we might further
complicate by the introduction of different mass fields for different connections, $\hat\psi^{(n)}_{U_{\uu_n}}$, or,
even more difficult, we may also introduce functions of such homogeneous expressions as
$\hat\psi^{\ }_{U_\uu}\hat\psi^{\,\dagger}_{U_\uu}/(U_\uu,U_\uu)_+$, \emph{etc.}, provided the resulting VEVs are
Lorentz and $U(1)$--gauge invariant.
All such expressions construct a complex of Dirac spinor and Maxwell bivector fields for a given Dirac spinor test
function, which may be tuned in very many ways to attempt to achieve a degree of empirical usefulness.
It is troubling that the huge wealth of high energy experimental data has been fitted so thoroughly into the conceptual
space of interacting Lagrangian and Hamiltonian formalisms that it is relatively difficult to apply that data as
restrictive principles in the conceptual space of the nonlinear test function formalism that is introduced here, beyond
the application of Lorentz and $U(1)$--gauge invariance and covariance and the possibly tenuous requirement that homogeneity
is preserved.

The choice of $\uu[U]$ determines a submanifold of the space of Dirac spinor test functions because of the invariance of
$U\mapsto U_\uu$ under the action of the $U(1)$ gauge transformation $U(x)\mapsto \rme^{-\rmi\theta(x)}U(x)$.
A geometric way to characterize the submanifold is first to observe that for a given Dirac spinor $U$ the 4--vectors
$J^\mu[U]=\overline{U}\gamma^\mu U$, $Z^\mu[U]=\overline{U}\gamma^5\gamma^\mu U$, and the real and imaginary parts
$X^\mu[U]$ and $Y^\mu[U]$ of $\overline{U}\gamma^\mu U^c$ constitute an orthogonal tetrad that is orthonormal up to a
constant multiple,
\begin{eqnarray*}
  J^\mu[U]J_\mu[U]&=&-Z^\mu[U]Z_\mu[U]=-X^\mu[U]X_\mu[U]=-Y^\mu[U]Y_\mu[U]\\
                  &=&[\overline{U}U]^2+[\overline{U}\rmi\gamma^5 U]^2.
\end{eqnarray*}
For a test function $U(x)$, $J^\mu[U_\uu(x)]=J^\mu[U(x)]$, $Z^\mu[U_\uu(x)]=Z^\mu[U(x)]$,
 $\sigma[U_\uu(x)]=\overline{U_\uu(x)}U_\uu(x)=\overline{U(x)}U(x)$, and
 $\omega[U_\uu(x)]=\overline{U_\uu(x)}\rmi\gamma^5 U_\uu(x)=\overline{U(x)}\rmi\gamma^5 U(x)$ are invariant under $U(1)$--gauge
transformations, however the $U(1)$--gauge invariant directions of the real and imaginary parts of
$\overline{U_\uu(x)}\gamma^\mu U_\uu^c(x)$ are in a fixed relation to spatial derivatives of $U(x)$ in the backward
light--cone of $x$ (given the choice of $\Xi^\mu=\partial^\mu G_{\mathrm{ret}}$ and of $\uu_\mu[U]$), independently
of the non--$U(1)$--gauge invariant directions of the real and imaginary parts of $\overline{U(x)}\gamma^\mu U^c(x)$.
Because of Fierz identities, the expression for $\uu[U]$ given above can be rewritten,
\begin{eqnarray*}
  \uu_\alpha[U](x)&=&\frac{\rmi}{2}\,
      \frac{J^\mu[U]\left[\overline{U}\gamma_\mu\partial_\alpha U-\overline{\partial_\alpha U}\gamma_\mu U\right]}
           {J^\mu[U]J_\mu[U]}\cr
   &=&\frac{\rmi}{2}\!
      \left[\frac{\sigma[U]}{(\sigma[U])^2+(\omega[U])^2}\overline{U}\partial_\alpha U
           +\frac{\omega[U]}{(\sigma[U])^2+(\omega[U])^2}\overline{U}\rmi\gamma^5\partial_\alpha U
                -\mathrm{c.c.}\right],
\end{eqnarray*}
so that the directions of the real and imaginary parts of $\overline{U_\uu(x)}\gamma^\mu U_\uu^c(x)$ are determined by the
imaginary parts of the spatial derivatives of $U$ relative to the projections $\overline{U}\partial_\alpha U$ and
$\overline{U}\rmi\gamma^5\partial_\alpha U$, weighted by functions in $\sigma[U]$ and $\omega[U]$.

\subsection{4--momentum of prepared states, interactions, and locality}\label{InteractionsLocality}
The underlying free quantum field structure has not been modified, so that the 4--momentum operator has the same expression in
terms of the free field creation and annihilation operators as it has for the free quantum fields $\emF$ and $\hat\psi$ and the
expected 4--momentum for a given state will still be in the forward light--cone.
The 4--momentum associated with a given set of test functions is different, however, because of the nonlinearity of the
quantum field operators as functionals of the test functions, so that the 4--momentum becomes a nontrivial function of the
wave--number components of the test functions.

The nonlinear modification we have introduced results in the possibility of interactions because of the modification
$U\mapsto U_\uu$, as well as there being a modification of the VEVs because of the introduction of the Maxwell bivector field.
For an example four--point VEV for the quantized free Dirac spinor field, smeared by test functions, we obtain
\[\VEV{\hat\psi^\dagger_{V_1}\hat\psi^\dagger_{V_2}\hat\psi^{\ }_{U_2}\hat\psi^{\ }_{U_1}}
             =\left|\begin{array}{c c}(V_{1},U_{1})_+\,&\,(V_{1},U_{2})_+\\
                                      (V_{2},U_{1})_+\,&\,(V_{2},U_{2})_+\end{array}\right|,
\]
for which we obtain a non--zero probability amplitude only if the determinant of the matrix here is non--zero.
In contrast, we obtain for the deformed VEV
$\VEV{\hat\Psi^\dagger_{V_1}\hat\Psi^\dagger_{V_2}\hat\Psi^{\ }_{U_2}\hat\Psi^{\ }_{U_1}}$,
\begin{eqnarray}
          &&\left|\begin{array}{c c}(V_{1\uu},U_{1\uu})_+\,&\,(V_{1\uu},U_{2\uu})_+\\
                                  (V_{2\uu},U_{1\uu})_+\,&\,(V_{2\uu},U_{2\uu})_+\end{array}\right|
               \VEV{\rme^{-\rmi\lambda\emF_{\!\Intd\uu[V_1]}}\rme^{-\rmi\lambda\emF_{\!\Intd\uu[V_2]}}
                    \rme^{\rmi\lambda\emF_{\!\Intd\uu[U_2]}}\rme^{\rmi\lambda\emF_{\!\Intd\uu[U_1]}}}\cr
\vspace{-3pt}\cr
          &=&\left|\begin{array}{c c}(V_{1\uu},U_{1\uu})_+\,&\,(V_{1\uu},U_{2\uu})_+\\
                                  (V_{2\uu},U_{1\uu})_+\,&\,(V_{2\uu},U_{2\uu})_+\end{array}\right|
          \exp\!\Big[\!-\!\SmallFrac{\lambda^2}{2}\!(\Intd\uu[V_1],\Intd\uu[V_1])_0
                     \!-\!\SmallFrac{\lambda^2}{2}(\Intd\uu[V_2],\Intd\uu[V_2])_0\cr
          &&\hspace{13.5em}
                     -\SmallFrac{\lambda^2}{2}(\Intd\uu[U_2],\Intd\uu[U_2])_0
                     \!-\!\SmallFrac{\lambda^2}{2}(\Intd\uu[U_1],\Intd\uu[U_1])_0\Big]\cr
          &&\qquad\times\rme^{-\lambda^2(\Intd\uu[V_1],\Intd\uu[V_2])_0
                              +\lambda^2(\Intd\uu[V_1]+\Intd\uu[V_2],\Intd\uu[U_1]+\Intd\uu[U_2])_0
                              -\lambda^2(\Intd\uu[U_2],\Intd\uu[U_1])_0},
            \label{FourPointVEV}
\end{eqnarray}
where we have used a Baker--Campbell--Hausdorff formula for the final step.
Explicitly, there are interactions in this construction because the determinant
$\left|\begin{array}{c c}(V_{1\uu},U_{1\uu})_+\,&\,(V_{1\uu},U_{2\uu})_+\\
                         (V_{2\uu},U_{1\uu})_+\,&\,(V_{2\uu},U_{2\uu})_+\end{array}\right|$ may be non--zero when the free
field probability amplitude is zero.
For the corresponding transition probability for $U_1,U_2\longrightarrow V_1,V_2$, we obtain
\begin{eqnarray*}
  &&\frac{\VEV{\hat\Psi^\dagger_{V_1}\hat\Psi^\dagger_{V_2}\hat\Psi^{\ }_{U_2}\hat\Psi^{\ }_{U_1}}
          \VEV{\hat\Psi^\dagger_{U_1}\hat\Psi^\dagger_{U_2}\hat\Psi^{\ }_{V_2}\hat\Psi^{\ }_{V_1}}}
         {\VEV{\hat\Psi^\dagger_{V_1}\hat\Psi^\dagger_{V_2}\hat\Psi^{\ }_{V_2}\hat\Psi^{\ }_{V_1}}
          \VEV{\hat\Psi^\dagger_{U_1}\hat\Psi^\dagger_{U_2}\hat\Psi^{\ }_{U_2}\hat\Psi^{\ }_{U_1}}}\cr
  &=&\quad\frac{\left|\begin{array}{c c}(V_{1\uu},U_{1\uu})_+\,&\,(V_{1\uu},U_{2\uu})_+\\
                                  (V_{2\uu},U_{1\uu})_+\,&\,(V_{2\uu},U_{2\uu})_+\end{array}\right|
          \left|\begin{array}{c c}(U_{1\uu},V_{1\uu})_+\,&\,(U_{1\uu},V_{2\uu})_+\\
                                  (U_{2\uu},V_{1\uu})_+\,&\,(U_{2\uu},V_{2\uu})_+\end{array}\right|}
         {\left|\begin{array}{c c}(V_{1\uu},V_{1\uu})_+\,&\,(V_{1\uu},V_{2\uu})_+\\
                                  (V_{2\uu},V_{1\uu})_+\,&\,(V_{2\uu},V_{2\uu})_+\end{array}\right|
          \left|\begin{array}{c c}(U_{1\uu},U_{1\uu})_+\,&\,(U_{1\uu},U_{2\uu})_+\\
                                  (U_{2\uu},U_{1\uu})_+\,&\,(U_{2\uu},U_{2\uu})_+\end{array}\right|}\cr
  &&\hspace{3em}\times\rme^{-\lambda^2(\Intd\uu[V_1]+\Intd\uu[V_2]-\Intd\uu[U_1]-\Intd\uu[U_2],
                                \Intd\uu[V_1]+\Intd\uu[V_2]-\Intd\uu[U_1]-\Intd\uu[U_2])_0}.
\end{eqnarray*}
We note that the 1--point factors $\exp[-\scriptHalf\lambda^2(\Intd\uu[U_1],\Intd\uu[U_1])_0]$, \emph{etc.}, are generally
insignificant, not only in this particular transition probability but also when we construct superpositions, insofar as we
may modify $U_1$ \emph{etc.} by a constant factor $U_1\mapsto \exp[\scriptHalf\lambda^2(\Intd\uu[U_1],\Intd\uu[U_1])_0]\,U_1$,
which eliminates the 1--point factors because $\Intd\uu[U_1]$ is invariant under multiplication by a real constant.
Such a modification may be automated by defining
$\hat\Psi^{'}_U=\exp[\scriptHalf\lambda^2(\Intd\uu[U],\Intd\uu[U])_0]\hat\Psi^{\ }_U=\,:\!\!\hat\Psi^{\ }_U\!\!:$, the same as the
effect of normal--ordering.
We also note that this transition probability and the next may be tuned substantially by replacing the exponential
$\rme^{\rmi\lambda\emF_{\!\Intd\uu[U]}}$ by an arbitrary function of $\emF_{\!\Intd\uu[U]}$, by using the constructions in
subsection \ref{OtherConstructions}, or by using the alternative $U(1)$--gauge connections and propagators in Appendices
\ref{GaugeConnections} and \ref{GaugePropagators}.

From a foundational point of view, we might assert that all observables should be constructed using only $\hat\Psi$, insofar
as all Maxwell bivector field phenomena are caused by the presence of nontrivial Dirac spinor fields, however we can use the
quantized free Maxwell bivector field $\emF(x)$ at least for practical purposes, with linear smearing by a bivector test function
$\fkf(x)$, say, to obtain the operator $\emF^{\ }_{\!\fkf}$, in which case we can construct nontrivial three--point
VEVs such as
\begin{equation}
  \VEV{\emF^\dagger_{\!\fkf}\hat\Psi^{'\,\dagger}_V\hat\Psi^{'}_U}=(V_\uu,U_\uu)_+
                                   \rmi\lambda\bigl[(\fkf,\Intd\uu[U])_0-(\fkf,\Intd\uu[V])_0\bigr]
     \rme^{\lambda^2(\Intd\uu[V],\Intd\uu[U])_0},\label{ThreePointVEV}
\end{equation}
resulting in the transition probability for $U\longrightarrow V,\fkf$,
\begin{eqnarray*}
  &&\frac{|(V_\uu,U_\uu)_+|^2}{(V_\uu,V_\uu)_+(U_\uu,U_\uu)_+}
      \frac{\lambda^2|(f,\Intd\uu[U])_0-(f,\Intd\uu[V])_0|^2}{(f,f)_0+\lambda^2|(f,\Intd\uu[V])_0|^2}\cr
  &&\hspace{12em}\times\rme^{-\lambda^2(\Intd\uu[V]-\Intd\uu[U],\Intd\uu[V]-\Intd\uu[U])_0},
\end{eqnarray*}
or, for the transition probability for $U,\mbox{\small anti-particle}(V)\longrightarrow \fkf$,
\begin{eqnarray*}
  &&\frac{|(V_\uu,U_\uu)_+|^2}{\left|\begin{array}{c c}(U_\uu,U_\uu)_+\,&\,(U_\uu,V^c_\uu)_+\\
                                                       (V^c_\uu,U_\uu)_+\,&\,(V^c_\uu,V^c_\uu)_+\end{array}\right|}
      \frac{\lambda^2|(f,\Intd\uu[U])_0-(f,\Intd\uu[V])_0|^2}{(f,f)_0}\cr
  &&\hspace{12.25em}\times\rme^{-\lambda^2(\Intd\uu[V]-\Intd\uu[U],\Intd\uu[V]-\Intd\uu[U])_0},
\end{eqnarray*}
with it being worthwhile to recall that $(V^c_\uu,V^c_\uu)_+=(V_\uu,V_\uu)_-$.
These constructions are electromagnetically trivial when $U(x)$ and $V(x)$ have a single component, in which case $\Intd\uu[U]$
and $\Intd\uu[V]$ are trivial, so that when constructing models in this nonlinear test function formalism we will in general
expect to use multi--component Dirac spinor test functions for which $\Intd\uu[\cdot]$ is non--trivial.

The multi--component variation of an electromagnetically nontrivial Dirac spinor test function for the nonlinear case is
analogous to the lesser detail of the polarizations and spins of S--matrix in and out states, and equally may not be known,
making it necessary to extend summations over polarization and spin components to include averaging over components of
the multi--component variation.
For example, for a free electron field in quantum electrodynamics we would expect to specify a wave--number $k_\mu$ and a
spinor $u^{(1)}(f)$ (representing a polarization frame $f$, which is constrained by the free field Dirac equation
to be aligned with $k_\mu$), resulting in a test function $\rme^{-\rmi\sVip{k}{x}}u^{(1)}(f)$; interacting quantum electrodynamics
also requires an infra--red regularization that is determined by the spectrum of the detector sensitivities.
In contrast, for the minimal nontrivial fermion field example Eq. (\ref{FFexample}), which also specifies the surrounding
electromagnetic field, we would specify a wave--number $k_\mu$, two constant spinors, $u^{(1)}(f)$ and $u^{(2)}(f)$
(representing up and down spinors in a polarization frame), and a polarization wave--number $\ell_\mu$, resulting in a
base test function $\cos(\Vip{\ell}{x})\rme^{-\rmi\sVip{k}{x}}u^{(1)}(f)+\sin(\Vip{\ell}{x})u^{(1)}(f)$; as for interacting
quantum electrodynamics, there must be suitable long--range regularization.
It will be of interest if there is \emph{any} systematic choice of test functions that gives effective models of this class for
given experimental preparations and measurements, and for a given theory of $U(1)$--gauge connections and propagators,
however it will only be useful if there are enough benefits to justify the increased complexity.

The quantized Maxwell bivector field $\emF$ is a free field, used essentially for practical purposes, so that although there
are $\emF$--$\hat\Psi^\dagger$--$\hat\Psi$ interactions, there are no photon--photon interactions; as it is presented here,
a model that is constructed using only $\hat\Psi$, using a macroscopic number of Dirac spinors as sources of the
quantized Maxwell bivector field, would be taken to be more fundamental.
There is apparently no way to deform $\emF^{\ }_{\!\fkf}$ to include Dirac spinor components that preserves homogeneity and
that is as immediately natural as the way in which we have deformed $\hat\psi_{U_\uu}$ to construct $\hat\Psi_{U_\uu}$.

Locality is preserved by the construction given here insofar as the free quantum field structure underlies the nonlinear
construction, however for the connection $\uu[U]$ that is constructed using a Dirac spinor test function $U$ to exist,
$U(x)$ must be non--zero ---even if arbitrarily small--- everywhere.
Local observables are nonetheless possible as limits in appropriate operator norms after the GNS--construction of a
vacuum sector or other Hilbert space, by taking sequences of observables in which the everywhere non--zero test functions used
to construct each instance in the sequence approach arbitrarily close to zero outside a given region of space--time.

\subsection{A diagrammatic presentation}
\begin{figure}[htb]
  \includegraphics{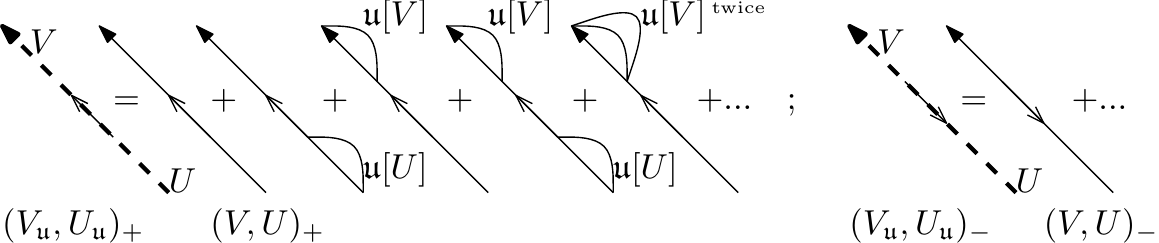}
\caption{\label{TwoPointxiDiagram}Diagrammatic presentation for $\VEV{\hat\xi_V^{\;\dagger}\hat\xi^{\ }_U\!}=(V^{\ }_\uu,U^{\ }_\uu)_+$
and for $\VEV{\hat\xi^{\ }_U\hat\xi_V^{\;\dagger}\!}=(V^{\ }_\uu,U^{\ }_\uu)_-$, showing both the particle--anti--particle
ordering, here $U\!\rightarrow\!V$ in both cases, and the operator ordering, $\hat\xi^{\ }_U$ acting on the vacuum vector
before or after $\hat\xi_V^{\;\dagger}$, resulting in $(V,U)_\pm$.
The labels $\uu[U]$ and $\uu[V]$ emphasize the essential feature that there is a massless propagation of $\uu[U]$ and $\uu[V]$
as well as the massive propagation of $U$ and $V$; although the propagation of $\uu[U]$ and $\uu[V]$ is not positive energy in
general, the aggregate propagation $(V^{\ }_\uu,U^{\ }_\uu)_\pm$, inherited from the quantized free Dirac spinor field, is
positive energy.}
\end{figure}

The algebraic presentation we have used is enough more straightforward than Feynman integrals, particularly because there
is no need to manage the combinatoric structure of renormalization, that a diagrammatic presentation is relatively superfluous,
however it is helpful to see what can be done.
We can write down an expansion for $\widetilde{U^{\ }_\uu}(k)$, although convergence will in general be slow,
\begin{eqnarray*}
  \widetilde{U^{\ }_\uu}(k)&=&\tilde{U}(k)
       +\rmi\!\int\!\tilde{U}(k')\Bigl[\widetilde{G_{\mathrm{ret}}}(k_1)k_1^\mu\widetilde{\uu_\mu[U]}(k_1)\Bigr]
               (2\pi)^4\delta^4(k'+k_1-k)\kInt{k'}\kInt{k_1}\cr
  &&   +\frac{\rmi^2}{2!}\!\int\!\tilde{U}(k')\Bigl[\widetilde{G_{\mathrm{ret}}}(k_1)k_1^\mu\widetilde{\uu_\mu[U]}(k_1)\Bigr]
                                                   \Bigl[\widetilde{G_{\mathrm{ret}}}(k_2)k_2^\mu\widetilde{\uu_\mu[U]}(k_2)\Bigr]\cr
  &&\hspace{7em}\times\quad               (2\pi)^4\delta^4(k'+k_1+k_2-k)\kInt{k'}\kInt{k_1}\kInt{k_2}+...,
\end{eqnarray*}
so that we can immediately draw a diagrammatic presentation for $\VEV{\hat\xi_V^{\;\dagger}\hat\xi^{\ }_U\!}=(V^{\ }_\uu,U^{\ }_\uu)_+$
and for $\VEV{\hat\xi^{\ }_U\hat\xi_V^{\;\dagger}\!}=(V^{\ }_\uu,U^{\ }_\uu)_-$, Fig. \ref{TwoPointxiDiagram}, showing both the
particle--anti--particle ordering and the operator ordering; there is no information about time--ordering, however, insofar as
time--ordering is a point-by-point operation that is relatively inappropriate for test functions $U$ and $V$ that are nowhere
zero, even if arbitrarily small almost everywhere.
Although the diagrams are presented as simply being added, powers of $\rmi$ are implied by the diagram rules.

\begin{figure}[htb]
  \includegraphics{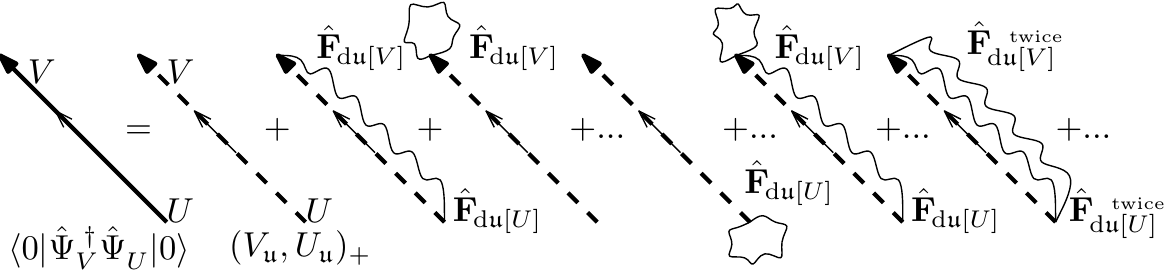}
\caption{\label{TwoPointPsiDiagram}Diagrammatic presentation for $\VEV{\hat\Psi_V^{\;\dagger}\hat\Psi^{\ }_U\!}$, using the diagram
for the $U(1)$--gauge invariant construction $(V^{\ }_\uu,U^{\ }_\uu)_+$ presented in Fig. \ref{TwoPointxiDiagram}.
The exponential factors $\exp[-\frac{\lambda^2}{2}(\Intd\uu[V],\Intd\uu[V])_0]$,
$\exp[-\frac{\lambda^2}{2}(\Intd\uu[V],\Intd\uu[U])_0]$, and $\exp[\lambda^2(\Intd\uu[V],\Intd\uu[U])_0]$ result in
arbitrary numbers of photon propagators from $V\!$ to $V\!$, from $U\!$ to $U$, and from $U\!$ to $V\!$.}
\end{figure}

\begin{figure}[htb]
  \includegraphics{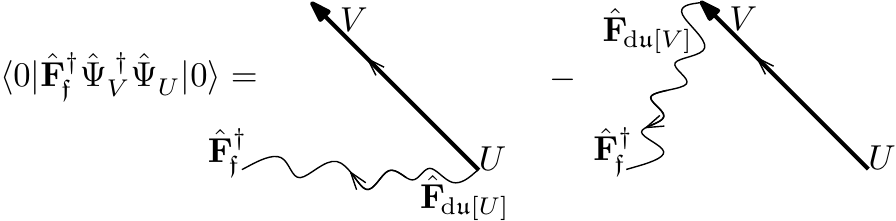}
\caption{\label{ThreePointDiagram}Diagrammatic presentation for $\VEV{\emF^\dagger_{\!\fkf}\hat\Psi^{\;\dagger}_V\hat\Psi^{\ }_U}$.
Note that the solid lines representing $\VEV{\hat\Psi_V^{\;\dagger}\hat\Psi^{\ }_U\!}$, \emph{etc.} already include photon
propagators from $U$ to $V$, \emph{etc.}, and all self--directed photon propagators.}
\end{figure}

\begin{figure}[htb]
  \includegraphics{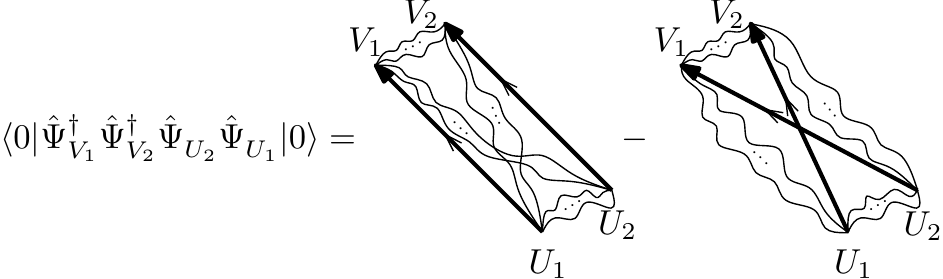}
  \caption{\label{FourPointDiagram}Diagrammatic presentation for
$\VEV{\hat\Psi^\dagger_{V_1}\hat\Psi^\dagger_{V_2}\hat\Psi^{\ }_{U_2}\hat\Psi^{\ }_{U_1}\!}$, representing exponentials in
$(\Intd\uu[V^{\ }_i],\Intd\uu[U^{\ }_j])_0$ as arbitrary multiples of photon propagators.
Labels $\emF^{\ }_{\!\Intd\uu[U_1]}$, \emph{etc.}, are omitted from the photon propagators adjacent to the test function labeled
$U_1$, \emph{etc.}}
\end{figure}

A diagrammatic presentation for $\VEV{\hat\Psi_V^{\;\dagger}\hat\Psi^{\ }_U\!}$, Fig. \ref{TwoPointPsiDiagram}, takes an
elementary form as a sum of $\VEV{\hat\xi_V^{\;\dagger}\hat\xi^{\ }_U\!}=(V^{\ }_\uu,U^{\ }_\uu)_+$ multiplied by zero or
more photon propagators, with the Maxwell bivector test functions provided by the $U(1)$--gauge curvatures $\Intd\uu[U]$ and
$\Intd\uu[V]$, including self--directed photons from $V$ to $V$ and from $U$ to $U$ and photons from $U$ to $V$.
$\VEV{\emF^\dagger_{\!\fkf}\hat\Psi^{\;\dagger}_V\hat\Psi^{\ }_U}$, derived in Eq. (\ref{ThreePointVEV}), modifies the compound diagram
element for $\VEV{\hat\Psi_V^{\;\dagger}\hat\Psi^{\ }_U\!}$ by the addition of photon lines $(\fkf,\Intd\uu[U])$ and
$(\fkf,\Intd\uu[V])$, which results in two diagrams, with opposite sign, in Fig. \ref{ThreePointDiagram}.
Finally, using the diagram elements we have constructed, and presenting exponentials in $(\Intd\uu[V],\Intd\uu[U])_0$ as
arbitrary multiples of photon propagators, a diagrammatic presentation for
$\VEV{\hat\Psi^\dagger_{V_1}\hat\Psi^\dagger_{V_2}\hat\Psi^{\ }_{U_2}\hat\Psi^{\ }_{U_1}\!}$, derived in Eq. (\ref{FourPointVEV}),
can be given in a moderately compact form, in Fig. \ref{FourPointDiagram}.

\subsection{Other homogeneity preserving deformations of Dirac spinor test functions}\label{OtherConstructions}
There is a wealth of higher degree constructions that use arbitrary powers of scalar invariants constructed using
$\delta\Intd\uu[U]$ and $\Intd\uu[U]$, which are both $U(1)$--gauge invariant and homogeneous of degree 0 over
the reals, so they may be used in many ways to construct deformations of $U_\uu$ that preserve $U(1)$--gauge
invariance and homogeneity.

For a given Dirac spinor test function, we can also construct a continuum of connections $\uu[U]$, as is shown in Appendix
\ref{GaugeConnections}.
Given any two such connections, $\uu_1[U]$ and $\uu_2[U]$, say, the difference $\uu_1[U]-\uu_2[U]$ is a $U(1)$--gauge
invariant vector field that is homogeneous of degree 0 over the reals, so we may use it anywhere we might use
$\delta\Intd\uu[U]$.

We also note the constructions of more general $U(1)$--gauge propagators in Appendix \ref{GaugePropagators}.

Even more than for the introduction of $\rme^{\rmi\lambda\emF_{\!\Intd\uu[U]}}$, however, there is a lack of decent
empirical or theoretical constraint, so we will not further pursue these and other constructions here.

\subsection{A $U(1)\!\times\!U(1)$--gauge connection}
It is possible also to construct a Dirac spinor test function that is invariant under the Lorentz invariant
$U(1)\!\times\!U(1)$--gauge transformation
\[  U(x)\mapsto \rme^{-\rmi(\theta_1(x)+\gamma^5\theta_2(x))}U(x),
\]
for which $\overline{U}\gamma^\mu U$ and $\overline{U}\gamma^5\gamma^\mu U$ are invariant; $\overline{U}\gamma^\mu U^c$
is invariant under the $\theta_2(x)$ component only; $\overline{U}U$, $\overline{U}\rmi\gamma^5 U$, and
$\overline{U}\gamma^{[\mu}\gamma^{\nu]}U$ are invariant under the $\theta_1(x)$ component only.
We introduce two connections, $\uu_\pm(x)$,
\[  \uu_{\pm\alpha}(x)=\frac{\rmi}{2}\,
      \left[\frac{\overline{U(x)}\gamma^\mu\Half(1\mp\gamma^5)U(x)\overline{U(x)}\gamma_\mu\Half(1\pm\gamma^5)\partial_\alpha U(x)}
                 {\overline{U(x)}\gamma^\mu\Half(1+\gamma^5)U(x)\overline{U(x)}\gamma_\mu\Half(1-\gamma^5) U(x)}-\mathrm{c.c.}\right],
\]
for which, under the $U(1)\!\times\!U(1)$--gauge transformation given above,
\[  \uu_{\pm\alpha}(x)\mapsto \uu_{\pm\alpha}(x)+\partial_\alpha\theta_1(x)\pm\partial_\alpha\theta_2(x),
\]
which allows us to construct the $U(1)\!\times\!U(1)$--gauge invariant
\begin{eqnarray*}
  U_{\underline{\uu}}&=&\exp\!\Big[\rmi\!\int\!\Bigl(\Xi_{+}^\mu(x-y)\scriptHalf(1+\gamma^5)\uu_{+\mu}[U](y)\cr
                     &&\hspace{4em}      +\;\Xi_{-}^\mu(x-y)\scriptHalf(1-\gamma^5)\uu_{-\mu}[U](y)\Bigr)
                                                 \mathrm{d}^4y\Big]U(x).
\end{eqnarray*}
This construction is somewhat more constrained than the construction of $U(1)$--gauge connections, but here again we may
replace $\Xi_\pm^\mu(x)$ by the constructions given in Appendix \ref{GaugePropagators} (with the constraint that if
$\Xi_\pm^\mu(x)$ are Dirac algebra--valued, they must be even--graded, $[\Xi_\pm^\mu(x),\gamma^5]=0$).

\section{Discussion}
Allowing nonlinearity as proposed here allows considerable arbitrariness, which has been somewhat controlled
by everywhere insisting upon Lorentz invariance, $U(1)$--gauge invariance, and homogeneity.
The connection with existing theory and experiment needs improvement, but the more elaborate structures of the
electroweak model and of quantum chromodynamics do not easily provide empirically principled constraints.

The constructive proposal in Section \ref{GTDS} can be seen as radical ---even though the adherence to Lorentz and
translation invariance and to non--supersymmetric quantum field theory on Minkowski space--time is intended to be
conservative--- and may contain missteps, however the analysis in Section \ref{NLQF} of the determination of the
renormalization scale appropriate for a model of a given experiment suggests an approximate general form for any
weakly nonlinear test function formalism: there should be a dependence on wave--numbers that are relevant to the
experiment, but no dependence on amplitude.
The amplitude invariant $U(1)$--gauge invariance of Section \ref{GTDS} is a more--or--less plausible response to
the analysis in Section \ref{NLQF}, given the general empirical success of gauge invariance, however this paper is
more a preliminary discussion than it is exhaustive and empirically flawless.
Even if the specific details of the constructions pursued here are not useful, it is nonetheless of interest that a
weakening of linearity that is no stronger than can be motivated by a reasonable reconsideration of renormalization
creates a possibility of modeling some kinds of interactions in a well--defined almost--Wightman field theory.

It is partly encouraging that very many models exist in a nonlinear test function formalism, whereas the existence of
renormalized interacting Lagrangian and Hamiltonian models is relatively delicate, however a nonlinear test function
formalism may be understood to be less explanatory than Lagrangian or Hamiltonian formalisms precisely because there
are so many models as well as because there is a relatively greater distance from classical dynamics, instead having
a closer relationship to signal analysis.
In contrast, we are very familiar with the quantization of Lagrangians and Hamiltonians and with symmetry, renormalization,
and other constraints, we have a great deal of experience in their effective use, and there is a comfortable if partly
illusory closeness to classical models.
Nonetheless it is to be hoped that it will be possible either to construct an alternative, signal analysis--oriented
formalism that is at least marginally empirically useful or to clearly state why such formalisms cannot be useful as
empirical, engineering--oriented models, and that in either case an understanding of whatever relationships there may
be with renormalized Lagrangian and Hamiltonian formalisms will be useful.

\appendix
\section{The quantized Dirac spinor field}\label{DiracAppendix}
From Itzykson and Zuber~\cite[\S 3-3]{IZ} we have a definition of the quantized Dirac spinor field,
\[
  \hat\psi(x)=\int\frac{\Intd^3\mathbf{k}}{(2\pi)^3}\frac{m}{k_0}
       \sum_{\alpha=1,2}\left[b_\alpha(k) u^{(\alpha)}(k)\rme^{-\rmi k\cdot x}+
                              d_\alpha^\dagger(k) v^{(\alpha)}(k)\rme^{\rmi k\cdot x}\right],
\]
where $u^{(\alpha)}(k)\rme^{-\rmi k\cdot x}$ and $v^{(\alpha)}(k)\rme^{\rmi k\cdot x}$ are two pairs of orthogonal solutions of
the Dirac equation $(i\gamma^\mu\partial_\mu-m)U(x)=0$, for which, from~\cite[\S 2-2-1]{IZ},
\begin{eqnarray*}
  & (k_\mu\gamma^\mu-m)u^{(\alpha)}(k)=0,\qquad (k_\mu\gamma^\mu+m)v^{(\alpha)}(k)=0,&\cr\vspace{2pt}\cr
  & \overline{u^{(\alpha)}(k)}u^{(\beta)}(k)=\delta_{\alpha\beta},\qquad
    \overline{v^{(\alpha)}(k)}v^{(\beta)}(k)=-\delta_{\alpha\beta},\qquad
    \overline{u^{(\alpha)}(k)}v^{(\beta)}(k)=0,&\cr\vspace{2pt}\cr
  & \sum_{\alpha=1,2}u^{(\alpha)}(k)\overline{u^{(\alpha)}(k)}=\frac{(k_\mu\gamma^\mu+m)}{2m},\qquad
    \sum_{\alpha=1,2}v^{(\alpha)}(k)\overline{v^{(\alpha)}(k)}=\frac{(k_\mu\gamma^\mu-m)}{2m}.&
\end{eqnarray*}
$k_0$ is the positive square root $\sqrt{|\mathbf{k}|^2+m^2}$, and $k_\mu=(k_0,\mathbf{k})$.
The operators $b_\alpha(k)$, $b_\alpha^\dagger(k)$, and $d_\alpha(k)$, $d_\alpha^\dagger(k)$ satisfy trivial
anti--commutation relations except for the non--manifestly Lorentz covariant expressions~\cite[Eq. 3-161]{IZ}
\begin{eqnarray*}
  \{b_\alpha(k),b_\beta^\dagger(q)\}&=&(2\pi)^3\frac{k_0}{m}\delta^3(\mathbf{k}-\mathbf{q})\delta_{\alpha\beta},\\
  \{d_\alpha(k),d_\beta^\dagger(q)\}&=&(2\pi)^3\frac{k_0}{m}\delta^3(\mathbf{k}-\mathbf{q})\delta_{\alpha\beta}.
\end{eqnarray*}
The algebraic elements $\gamma^\mu$ generate the Clifford algebra $\mathcal{C}(1,3)$, satisfying the anti--commutation
relations $\gamma^\mu\gamma^\nu+\gamma^\nu\gamma^\mu=2g^{\mu\nu}$ with metric $\mathrm{diag}(1,-1,-1,-1)$; taken with
an imaginary $\rmi$ we obtain the Dirac algebra, which is isomorphic to the algebra of 4--dimensional matrices over
the complex field.
The Dirac conjugate is a complex anti--linear anti--automorphism of the Dirac algebra, satisfying
$\overline{\gamma^\mu}=\gamma^\mu$, $\overline{\rmi}=-\rmi$, $\overline{AB}=\overline{B}\,\overline{A}$,
which uniquely determines a Lorentz invariant Dirac conjugate of a Dirac spinor;
the charge conjugate is a complex anti--linear automorphism, satisfying $(\gamma^\mu)^c=-\gamma^\mu$, $\rmi^c=-\rmi$,
$(AB)^c=A^cB^c$, which determines a Lorentz invariant charge conjugate of a Dirac spinor only up to a complex phase;
a complex conjugate is definable for the Dirac algebra, $(\gamma^\mu)^*=\gamma^\mu$, $\rmi^*=-\rmi$,
$(AB)^*=A^*B^*$, but this does not determine a Lorentz invariant conjugate of a Dirac spinor.

To obtain a manifestly Lorentz covariant presentation of these anti--commutation relations, we may rescale the creation
and annihilation operators, taking $b'_\alpha(k)=2m\,2\pi\delta(\Vip{k}{k}-m^2)\theta(k_0)b_\alpha(k)$ and
$d'_\alpha(k)=2m\,2\pi\delta(\Vip{k}{k}-m^2)\theta(k_0)d_\alpha(k)$,
equivalent to rescaling by $m/k_0$ on the mass-shell, so that
\begin{eqnarray*}
  \hat\psi(x)&=&\int 2m\,2\pi\delta(\Vip{k}{k}-m^2)\theta(k_0)
       \sum_{\alpha=1,2}\left[b_\alpha(k) u^{(\alpha)}(k)\rme^{-\rmi k\cdot x}+
                              d_\alpha^\dagger(k) v^{(\alpha)}(k)\rme^{\rmi k\cdot x}\right]\frac{\Intd^4k}{(2\pi)^4}\cr
    &=&\int\sum_{\alpha=1,2}\left[b'_\alpha(k) u^{(\alpha)}(k)\rme^{-\rmi k\cdot x}+
                              {d'}_\alpha^\dagger(k) v^{(\alpha)}(k)\rme^{\rmi k\cdot x}\right]\frac{\Intd^4k}{(2\pi)^4}.
\end{eqnarray*}
We enforce the restriction to the forward mass--shell in the anti--commutation relations instead of in the definition of the
spinor field, so that, manifestly Lorentz covariantly,
\begin{eqnarray*}
  \{b'_\alpha(k),{b'}_\beta^\dagger(q)\}&=&2m(2\pi)^4\delta^4(k-q)\delta_{\alpha\beta}2\pi\delta(\Vip{k}{k}-m^2)\theta(k_0),\\
  \{d'_\alpha(k),{d'}_\beta^\dagger(q)\}&=&2m(2\pi)^4\delta^4(k-q)\delta_{\alpha\beta}2\pi\delta(\Vip{k}{k}-m^2)\theta(k_0).
\end{eqnarray*}
Whatever the scaling of the unobservable creation and annihilation operators, we have for $\hat\psi_\xi(x)$ the
Lorentz covariant anti--commutation relations~\cite[Eq.~3-170]{IZ}
\begin{eqnarray*}
  \{\hat\psi_\xi(x),\overline{\hat\psi_{\xi'}(x')}\}
             &=& \iS_{\xi\xi'}(x,x')=\left(\rmi\gamma^\mu\frac{\partial}{\partial x^\mu}+m\right)_{\xi\xi'}\iD_m(x-x')\cr
             &=&\int 2\pi\delta(\Vip{k}{k}-m^2)\varepsilon(k_0)(k_\mu\gamma^\mu+m)_{\xi\xi'}\rme^{-\rmi k\cdot(x-x')}
                     \frac{\Intd^4k}{(2\pi)^4}.
\end{eqnarray*}

We may put the above structure into an equivalent smeared operator form, smearing the Dirac spinor operator--valued
distribution $\hat\psi_\xi(x)$ with a Dirac spinor test function $U_\xi(x)$ to obtain a scalar operator that is a
linear functional of $U(x)$,
\[
  \hat\psi^{\ }_U=\int\overline{U^c(x)}\hat\psi(x)\Intd^4x=\dOP^{\ }_{U^c}+\bOP^{\;\dagger}_U,
\]
where $\bOP^{\ }_U$ and $\dOP^{\ }_U$ are complex anti--linear in $U$.
We have to introduce charge conjugation to ensure that $\hat\psi^{\ }_U$ follows the convention that it should be complex--linear in $U$,
and correspondingly for $\dOP_{U^c}$.
For these scalar smeared operators, we have the anti--commutation relations
\begin{eqnarray*}
  \{\hat\psi^{\ }_U,\hat\psi_V^{\;\dagger}\}=(V,U)&=&\{\bOP^{\;\dagger}_U,\bOP_V^{\ }\}+\{\dOP^{\ }_{U^c},\dOP^{\;\dagger}_{V^c}\}
                                                      =(V,U)_++(V,U)_-\cr
                           &=&\int\frac{\Intd^4k}{(2\pi)^4}2\pi\delta(\Vip{k}{k}-m^2)
                              \varepsilon(k_0)\overline{\tilde V(k)}(k_\mu\gamma^\mu+m)\tilde U(k)\\
   &=& \int \overline{V(x)}\iS(x,x')U(x')\Intd^4x\Intd^4x',\\
  \{\bOP^{\ }_U,\bOP_V^{\;\dagger}\}=\{\dOP^{\ }_U,\dOP_V^{\;\dagger}\}=(U,V)_+
                           &=&\int\frac{\Intd^4k}{(2\pi)^4}2\pi\delta(\Vip{k}{k}-m^2)
                              \theta(k_0)\overline{\tilde U(k)}(k_\mu\gamma^\mu+m)\tilde V(k)\cr
                           &=&\VEV{\hat\psi^{\;\dagger}_U\hat\psi^{\ }_V}.
\end{eqnarray*}
The arbitrary phase that appears to be introduced by charge conjugation in these equations in fact cancels because of the
identities $\overline{A^c}\gamma^\mu B^c=\overline{B}\gamma^\mu A$ and $\overline{A^c}\,B^c=-\overline{B}\,A$.
Explicitly, for $\{\dOP^{\ }_{U^c},\dOP^{\;\dagger}_{V^c}\}$, we have
\begin{eqnarray*}
  \{\dOP^{\ }_{U^c},\dOP^{\;\dagger}_{V^c}\}&=&\int\frac{\Intd^4k}{(2\pi)^4}2\pi\delta(\Vip{k}{k}-m^2)
                                      \theta(k_0)\overline{\widetilde{U^c}(k)}(k_\mu\gamma^\mu+m)\widetilde{V^c}(k),\cr
                                   &=&\int\frac{\Intd^4k}{(2\pi)^4}2\pi\delta(\Vip{k}{k}-m^2)
                                      \theta(k_0)\overline{\left[\widetilde{V^c}(k)\right]^c}(k_\mu\gamma^\mu-m)
                                                           \left[\widetilde{U^c}(k)\right]^c,\cr
                                   &=&\int\frac{\Intd^4k}{(2\pi)^4}2\pi\delta(\Vip{k}{k}-m^2)
                                      \theta(k_0)\overline{\tilde V(-k)}(k_\mu\gamma^\mu-m)\tilde U(-k),\cr
                                   &=&-\int\frac{\Intd^4k}{(2\pi)^4}2\pi\delta(\Vip{k}{k}-m^2)
                                      \theta(-k_0)\overline{\tilde V(k)}(k_\mu\gamma^\mu+m)\tilde U(k)\eqdef (V,U)_-\cr
                                   &=&\VEV{\hat\psi^{\ }_U\hat\psi^{\;\dagger}_V}.
\end{eqnarray*}

\section{Alternative $U(1)$--gauge connections}\label{GaugeConnections}
The main text suggests that the most obvious $U(1)$--gauge connection is
$$\uu_\alpha[U](x)=\frac{\rmi}{2}
      \left[\frac{\overline{U(x)}\gamma^\mu U(x)\overline{U(x)}\gamma_\mu\partial_\alpha U(x)}
                 {\overline{U(x)}\gamma^\mu U(x)\overline{U(x)}\gamma_\mu U(x)}-\mathrm{c.c.}\right].
$$
More generally, if we take $M_1[U]$ and $M_2[U]$ to be Lorentz invariant functions of $U$, and
\[
  \uu_\alpha[U](x)=\frac{\rmi}{2}
       \left[M_1[U]\frac{\overline{U(x)}\partial_\alpha U(x)}{\overline{U(x)}U(x)}
            +M_2[U]\frac{\overline{U(x)}\mathrm{i}\gamma^5\partial_\alpha U(x)}{\overline{U(x)}\mathrm{i}\gamma^5U(x)}
            -\mathrm{c.c.}\right],
\]
we obtain a $U(1)$--gauge connection if $\mathrm{Re}\bigl[M_1[U]+M_2[U]\bigr]=1$, with $\uu_\alpha[U](x)$ homogeneous of
degree zero over the reals if $M_1[U]$ and $M_2[U]$ are homogeneous of degree zero over the reals. 
Constructions that use other Lorentz invariants, such as $\overline{U(x)}\gamma^\mu U(x)\overline{U(x)}\gamma_\mu U(x)$,
all reduce to the same form by the application of Fierz identities such as
\begin{eqnarray*}
  \overline{U(x)}\gamma^\mu U(x)\overline{U(x)}\gamma_\mu U(x) &=& [\overline{U(x)}U(x)]^2+[\overline{U(x)}\mathrm{i}\gamma^5U(x)]^2,\\
  \overline{U(x)}\gamma^\mu U(x)\overline{U(x)}\gamma_\mu \partial_\alpha U(x) &=&
         \overline{U(x)}U(x)\overline{U(x)}\partial_\alpha U(x)
        +\overline{U(x)}\mathrm{i}\gamma^5U(x)\overline{U(x)}\mathrm{i}\gamma^5\partial_\alpha U(x).
\end{eqnarray*}
Also because of Fierz identities, $M_1[U]$ and $M_2[U]$ may both be taken to be functionals of $\overline{U(x)}U(x)$ and
$\overline{U(x)}\mathrm{i}\gamma^5U(x)\overline{U(x)}$.
We also note that the sum of a $U(1)$--gauge connection and a 4--vector is a $U(1)$--gauge connection, and that
the divergence $\delta\Intd\uu_\alpha[U]$ of the $U(1)$--gauge curvature $\Intd\uu_{\alpha\mu}[U]$ is a 4--vector,
so that we may add a weighted multiple of this and higher derivatives to obtain a new $U(1)$--gauge connection.

There is no reason why any such constructions must be of use in detailed physical models, but they are available
within the admittedly only partial empirical constraints adopted here of Lorentz invariance, $U(1)$--gauge invariance,
and homogeneity of degree 1 over the reals.

\section{Alternative $U(1)$--gauge propagators}\label{GaugePropagators}
The requirement that $\partial_\mu\Xi^\mu(x)=\delta^4(x)$ may be Lorentz covariantly satisfied in a number of ways.
The main text suggests the most obvious, the gradient of a retarded zero--mass Green's function in 3+1--dimensions,
$\Xi^\mu(x)=\partial^\mu G_{\mathrm{ret}}(x)$.
The next obvious alternative is an arbitrary affine combination of $\partial^\mu G_{\mathrm{ret}}(x)$ and
$\partial^\mu G_{\mathrm{adv}}(x)$, plus $\partial^\mu\xi(x)$, where $\xi(x)$ is a Lorentz invariant
solution of the wave equation, however lower--dimensional propagators may also be used when the direction of
the projection can be specified by the test function,
\[  U_\uu(x)=\exp{\!\left[\rmi\!\int\!\Xi[U(x)]^\mu(x-y)\uu_\mu[U](y)\mathrm{d}^4y\right]}U(x).
\]
It is essential that the function $\Xi[U(x)]^\mu$ is independent of the point $x-y$ for $U_\uu(x)$ to be $U(1)$--gauge invariant.
For example, a 4--vector distribution used by Steinmann~\cite{Steinmann} is $\Xi_0(x)=\Xi_1(x)=\Xi_2(x)=0$,
$\Xi_3(x)=\delta(x^0)\delta(x^1)\delta(x^2)\theta(x^3)$, which may be given in manifestly Lorentz covariant form as a
function of a Dirac spinor $Z$ that is non--zero only along a space--like half--line,
\begin{eqnarray*}
  \Xi[Z]^\mu(x)&=&-\left(\overline{Z}\gamma^\alpha Z\overline{Z}\gamma_\alpha Z\right) \overline{Z}\gamma^5\gamma^\mu Z
                      \delta(\overline{Z}\gamma_\alpha Z x^\alpha)\theta(\overline{Z}\gamma^5\gamma_\alpha Z x^\alpha) \cr
           &&\hspace{10em}\times\
                      \delta(\Re[\overline{Z^c}\gamma_\alpha Z] x^\alpha)\delta(\Im[\overline{Z^c}\gamma_\alpha Z] x^\alpha).
\end{eqnarray*}
Alternatively, in a slightly generalized form that is non--zero both along a space--like half--line and along a time--like half--line in
proportions specified by the scalar--pseudo--scalar phase of $Z$, we may use
\begin{eqnarray*}
  \Xi'[Z]^\mu(x)&=&(\overline{Z}Z)^2 \overline{Z}\gamma^\mu Z
                      \theta(\overline{Z}\gamma_\alpha Z x^\alpha)\delta(\overline{Z}\gamma^5\gamma_\alpha Z x^\alpha)
                      \delta(\Re[\overline{Z^c}\gamma_\alpha Z] x^\alpha)\delta(\Im[\overline{Z^c}\gamma_\alpha Z] x^\alpha)\cr
  &&\hspace{-3em}-(\overline{Z}\rmi\gamma^5 Z)^2 \overline{Z}\gamma^5\gamma^\mu Z
                      \delta(\overline{Z}\gamma_\alpha Z x^\alpha)\theta(\overline{Z}\gamma^5\gamma_\alpha Z x^\alpha)
                      \delta(\Re[\overline{Z^c}\gamma_\alpha Z] x^\alpha)\delta(\Im[\overline{Z^c}\gamma_\alpha Z] x^\alpha).
\end{eqnarray*}
To show that $\delta\Xi[Z](x)=\delta\Xi'[Z](x)=\delta^4(x)$ we use the equal--length Lorentzian orthogonality of the tetrad
$\overline{Z}\gamma^\mu Z$, $\overline{Z}\gamma^5\gamma^\mu Z$, $\Re[\overline{Z^c}\gamma^\mu Z]$, and
$\Im[\overline{Z^c}\gamma^\mu Z]$, and the identity
$(\overline{Z}Z)^2+(\overline{Z}\rmi\gamma^5 Z)^2=\overline{Z}\gamma^\alpha Z\overline{Z}\gamma_\alpha Z$.
For constructions that have a dependence on the point $x$ as well as on the separation $x-y$, the connection
$\uu_\alpha[U_\uu](x)$ in general will not be divergence--free.

As another example, we may use a 4-vector distribution that in elementary coordinates is $\Xi^0(x)=0$,
$\Xi^i(x)=\delta(x_0)\frac{x^i}{4\pi|\mathbf{x}|^3}$, or that in manifestly covariant form, for a time-like
4-vector $v$, is
\begin{equation}
  \Xi''[v]^\mu(x)=\delta(v\!\cdot\!x)\frac{[(v\!\cdot\!v)x^\mu-(v\!\cdot\!x)v^\mu]}
                                        {4\pi\left[\sqrt{(v\!\cdot\!x)^2-(v\!\cdot\!v)(x\!\cdot\!x)}\right]^3},
\end{equation}
We may use $\Xi[U(x)]^\mu(x-y)$, $\Xi'[U(x)]^\mu(x-y)$, $\Xi''[\overline{U(x)}\gamma U(x)]^\mu(x-y)$, or a convex sum of these
and similar constructions in the construction of $U_\uu$.
We note that all the above constructions are invariant under multiplication by a real--valued positive--definite scalar function
$E(x)$, $U(x)\mapsto E(x)U(x)$, as well as under $U(1)$--gauge transformations.

Finally, $\Xi^\mu(x)$ may be Dirac algebra--valued (noting, however, that $\partial_\mu\Xi^\mu(x)=\delta^4(x)$ is a scalar),
in which case we may use, for example,
\[
  \Xi^\mu(x)=\Xi^\mu_{(0)}(x)
      +\partial_\nu\!\left[\gamma^{[\mu}\partial^{\nu]}\phi^{\ }_{\!(1)}(x)
                          +\rmi\gamma^{[\mu}\gamma^{\nu]}\phi^{\ }_{\!(2)}(x)
                          +\gamma^5\gamma^{[\mu}\partial^{\nu]}\phi^{\ }_{\!(3)}(x)\right]\!
      +\rmi\gamma^5\partial^\mu\xi^{\ }_{(4)}(x).
\]
where $\phi^{\ }_{\!(1)}(x)$, $\phi^{\ }_{\!(2)}(x)$, and $\phi^{\ }_{\!(3)}(x)$ are arbitrary Lorentz invariant distributions,
possibly including massive components, $\Xi^\mu_{(0)}(x)$ is a scalar $U(1)$--gauge connection, and $\xi^{\ }_{(4)}(x)$ is a
Lorentz invariant solution of the wave equation.
With this construction, the connection $\uu_\alpha[U_\uu](x)$ in general will not be divergence--free; at the cost of still
less tractability, we may also introduce Dirac algebra--valued $U(1)$--gauge connections of the form $\Xi[U(x)]^\mu(x-y)$.

There is no reason why any such constructions must be of use in detailed physical models, but they are available
within the admittedly only partial empirical constraints adopted here of Lorentz invariance, $U(1)$--gauge invariance,
and homogeneity of degree 1 over the reals.


\begin{thebibliography}{}
\bibitem{Fredenhagen2007}
  {K. Fredenhagen, K.-H. Rehren, and E. Seiler, Quantum Field Theory: Where We Are,
                                                \textit{Lect. Notes Phys.} \textbf{721}, 61-87 (2007).}
\bibitem{Buchholz2000}
  {D. Buchholz and R. Haag, The quest for understanding in relativistic quantum physics,
                            \textit{J. Math. Phys.} \textbf{41}, 3674-3697 (2000).}
\bibitem{Haag}
  {R. Haag, \textit{Local Quantum Physics}, Springer, Berlin (1996).}
\bibitem{WilsonKogut}
  {K. G. Wilson and J. B. Kogut, The renormalization group and the $\varepsilon$ expansion,
                                 \textit{Phys. Rep.} \textbf{12}, 75-200 (1974).}
\bibitem{Rosten}
  {O. J. Rosten, Fundamentals of the exact renormalization group,
                 \textit{Phys. Rep.} \textbf{511}, 177-272 (2012).}
\bibitem{IZ}
  {C. Itzykson and J.-B. Zuber, \textit{Quantum Field Theory}, McGraw-Hill International Edition, Singapore (1985).}
\bibitem{Steinmann}
  {O. Steinmann, Perturbative QED in Terms of Gauge Invariant Fields, \textit{Ann. Phys.} \textbf{157}, 232-254 (1984).}
\end{thebibliography}
\end{document}